\documentclass[aps,twocolumn,amsmath,amssymb,floatfix]{revtex4}
\usepackage{graphicx}
\usepackage{amssymb}
\usepackage{color}
\usepackage{float}
\usepackage{mathrsfs}
\usepackage{epstopdf}
\usepackage{placeins}

\usepackage[normalem]{ulem}  

\renewcommand\sout{\bgroup \color{red} \ULdepth=-.5ex \ULset}

\begin{document}



\title{A geometrical interpretation of the Thomas theorem and the Efimov States}

\author{H. Zheng$^{1}$\footnote{Email address: zhengh@snnu.edu.cn}, A. Bonasera$^{2,3}$}

\affiliation{$^{1}$ School of Physics and Information Technology, Shaanxi Normal University, Xi'an 710119,  China}

\affiliation{$^{2}$ Cyclotron Institute, Texas A\&M University, College Station, TX 77843, USA}

\affiliation{$^{3}$ Laboratori Nazionali del Sud, INFN, via Santa Sofia, 62, 95123 Catania, Italy}


\begin{abstract}
Using a generalized Bohr model and the hyper-spherical formalism for a three-body system, we derive the Thomas theorem assuming a simple interaction depending on the range of the potential.  We discuss the conditions for which an unbound two-body system produces a bound three-body system and derive universal energy functions. We apply our model to $^{4}$He and Triton atoms as well as to the triton nucleus. Using their scattering lengths and effective ranges, we are able to reproduce the two-body or the three-body binding energies with only one parameter fitted. Prediction for excited (Efimov) levels are also given and in particular we demonstrate that for some hyper-angles two equal minima appear which indicate a phase (shape) transition similar to the Landau's theory of phase transition. We suggest that the observed excited levels in two different experiments for the triton nucleus are indeed Efimov levels and there may be more surprises.

\end{abstract}


\maketitle

In 1935, L.H. Thomas \cite{Thomas:1935zz} demonstrated that if a {\it quantum} two-body system has at least one loosely bound state then a system made of three particles becomes strongly bound. In particular he showed in the limit that the two body interaction becomes a delta function (i.e., its range $r_0\rightarrow 0$), the corresponding three-body system is unbound from below, i.e., its binding energy becomes $-\infty$.  In 1970 Efimov \cite{Efimov:1970zz}, building up on the Thomas effect, showed that if the scattering length $a\gg r_0$, then the three-body system displays a geometrical series of excited levels roughly spaced in the interval $\frac{\hbar^2}{mr_0^2}$ and $\frac{\hbar^2}{ma^2}$. Thus, if $r_0\rightarrow 0$ and $a\rightarrow \infty$, an infinite series of excited levels appears. In particular, the number of excited three-body levels, is roughly given by the simple relation  \cite{Efimov:1970zz}:
\begin{equation}
N\rightarrow \frac{s_0}{\pi}\ln(|a|/r_0), \label{eq1}
\end{equation} 
where $s_0=1.00624$ is a universal constant. We stressed the fact above that a {\it quantum} system shows these properties, i.e., there is no classical analog to the Thomas and Efimov effects. Originally, these properties were discussed for nuclear systems in particular to derive the binding energy of t and $^{3}$He nuclei \cite{Thomas:1935zz} as well as the first excited level of $^{12}$C, the Hoyle state, which was suggested to be an Efimov level \cite{Efimov:1970zz}, see Ref. \cite{Braaten:2006vd} for more details. The deuterium nucleus is a loosely bound system, $E_d=2.225$ MeV, and the nucleon-nucleon scattering length in the triplet state $a_t=5.42$ fm while the effective range $r_t=1.76$ fm \cite{Braaten:2006vd}.  These values are not very large and Eq. (\ref{eq1}) would support at most one excited (Efimov) level for the three body system, while the ground state of deuterium should be of the order of $\frac{\hbar^2}{ma^2}=1.4$ MeV, i.e., close to the experimental value. The three-body nuclei, t and $^{3}$He are much more bound than the d-nucleus, $E_t$=8.48 MeV and $E_{^{3}He}$=7.72 MeV  in agreement to the Thomas theorem  \cite{Thomas:1935zz}. Until recently, no excited levels of these nuclei were known, but two independent experiments \cite{Rogachev:2003mv, russ} have found some anomalies in the energy spectrum of t nuclei, which have been interpreted as an excited level of t at about $E_t^* \approx 7.0$ MeV  in Ref. \cite{russ} or final state interaction in Ref. \cite{Rogachev:2003mv}. 
The derivation of the Thomas (TS) and Efimov states does not involve any particular physical system, just that the two-body interaction is short range. Thus we are dealing with universal properties and we can apply the same to other systems, for instance atoms. In this case we can have a large variety of natural scattering and effective ranges for Fermion and Boson systems. Furthermore, using magnets or other devices we can effectively change the scattering lengths to almost any value using for instance the Feshbach resonances \cite{Braaten:2006vd}. In this framework, the TS and the ES have been experimentally demonstrated in recent years for a variety of atomic systems \cite{zaccanti:nature09, greenephytoday10, Kunitski:2015qth, efo1, efo2, efo3, he4cal} and most criticism has been put definitely to bed.

The quantal and universal features of the TS and ES prompted us to investigate a simple geometric model to help our physical intuition and possibly explore new venues. In exchange we had to sacrifice some of the mathematical beauty, which has been derived in the past 80 years or so \cite{Thomas:1935zz, Efimov:1970zz, Braaten:2006vd, greenephytoday10}. At the heart of our approach is the Bohr model \cite{bohr}, which gives a reasonable description of many systems' ground and excited states. In order to greatly simplify the math and derive universal expressions, we have used hyper-angles to describe the three-body system \cite{Braaten:2006vd}. Within this framework, we first derive the TS result and go beyond it by investigating how `loosely' bound the two-body system needs to be before the three-body one becomes unbound. We later apply the model to some physical examples from atomic and nuclear physics. We show that by simply fitting the interaction strength to the two-body ground state (when known) gives a very good prediction for the three-body system (when known) or vice versa. Ground states, or TS, are found when the three hyper-angles are equal, which corresponds to an equilater triangular geometric distribution. Excited levels are also found and the ES corresponds to one small hyperangle (less than $\approx \frac{\pi}{12}$) and the other two approaching $\frac{\pi}{3}$. Geometrically this corresponds to a monomer plus a dimer system (1+2). This is, in our opinion, an experimental signature of an ES, like for instance the $^{12}$C Hoyle state (HS) which decays mostly into the $^{8}$Be(gs)+$\alpha$ \cite{Efimov:1970zz, Zheng:2018iml} or $^{4}$He atoms \cite{Kunitski:2015qth}.  Recent experimental results of the HS decay give an upper value for the ratio of the direct decay (DD) to sequential decay (SD) less than 0.043\% \cite{Raduta:2011yz, Manfredi:2012zz, DellAquila:2017ppe, Smith:2017jub, Kirsebom:view2017, Kirsebom:2012zza, Itoh:2014mwa, Freer:1994zz, Zhang:2018tzn}. For an ES we expect the probability to decay into different channels to be exactly zero, thus the need for higher experimental precision \cite{Zheng:2018iml}. For atomic systems, the situation is more advanced respect to nuclei for the reason discussed above and our definition of ES may fit well the experimental results \cite{Braaten:2006vd, efimov:nature09, zaccanti:nature09, Kunitski:2015qth}. However, distinctions must be done when the scattering length is positive (which admits two-body bound states) or negative (no bound states, resonances at most in the two-body channel). Triton ($^{3}$He) nuclei and $^{4}$He atoms are examples of the first case while $^{8}$Be and polarized triton atoms might be examples of the unbound two body systems. These examples involve Bosons and non-identical Fermions, and in some cases the Coulomb interaction comes into play. This can be easily incorporated in the model and explicitly breaks the universality of the functions we derive below. For the goals of this work, we will not discuss them any further and concentrate on systems not subject to long-range interactions. The latter are interesting on their own and we will further explore them in a following publication by treating the long range force as an external field similar to the Landau's theory of phase transitions \cite{landau, huang}. 

 A novel result from this approach is the transition from the ground state (TS) to an excited (ES) level.  For positive scattering lengths and comparable effective lengths we find a {\it critical}  hyper-angle for which two energy minima exist for the trimer. A comparison to phase transitions can be drawn and in particular we can write the Landau's (free) energy for phase transitions in terms of one hyper-angle (which plays the role of the `temperature') and the hyper-distance (which plays the role of the order parameter, for instance the density in a liquid-gas phase transition).  In particular if the effective and scattering lengths are comparable, then the `phase transition' is first order (similar to the isotropic-nematic transition \cite{revlandau}). When the effective length is negligible respect to the scattering length, the two minima disappear and the `phase transition' becomes second order. Large negative scattering lengths display bound states but for a region in hyper-angles (near $\frac{\pi}{8}-\frac{\pi}{10}$) unbound resonances appear which become bound states again (ES) for smaller hyper-angles.  The concept of phase transition in this context might be associated to a shape transition or a transition from TS to ES. It is quite impressive to have so many features in a quantum system made of three particles only!
 
 The energy of a N-particles with equal mass system interacting through a two-body potential can be written as:
 \begin{eqnarray}
 E&=&\sum_{i=1}^{N}\frac{p_i^2}{2m}+\sum_{i=1<j}^{N}V(r_{ij})\nonumber\\
 &=&\frac{2}{N}\sum_{i=1<j}^{N}\frac{p_{ij}^2}{2\mu}+\sum_{i=1<j}^{N}V(r_{ij}).
 \end{eqnarray}
Where ${\bf p_{ij}=\frac{p_i-p_j}{2}}$ is the relative momentum of particle $i$ and particle $j$, $\mu=\frac{m}{2}$ is the reduced mass and ${\bf r_{ij}=r_i-r_j}$. Adopting the Bohr Model, we can write $p_{ij} r_{ij}=n \hbar$, and we will restrict our considerations to n=1, i.e., the system's ground state. We will make a further assumption and write the short-range two-body potential as:
\begin{equation}
V(r_{ij})= -c\frac{\hbar^2}{ma^2} \exp[-(r_{ij}-r_0)^2/a^2].\label{pv}
\end{equation}
Where $c$ is the (only) fitting parameter. Using this form (or any other short range potential form which can be written as $V(r_{ij}, r_0, a)$) of the two-body potential and the Heisenberg relation above, we can write the energy of a two-body system in a-dimensional form:
\begin{eqnarray}
E_2&=&\frac{\hbar^2}{mr_{ij}^2}+V(r_{ij})\nonumber\\
&=&\frac{\hbar^2}{ma^2}[\frac{1}{x^2}-ce^{-(x-x_s)^2}]\nonumber\\
&=&\frac{\hbar^2}{ma^2}[\frac{1}{x^2}-ce^{-(x-x_s)^2}], \label{e2}
\end{eqnarray}
where $x=r_{ij}/a$ and $x_s=r_0/a$. The coefficient before the square parenthesis determines the systems' units. To deal with a three-body system, we introduce the hyper-angles $\alpha_k$ and hyper-radius R coordinates \cite{Braaten:2006vd}:
\begin{equation}
R^2=\frac{1}{3}(r_{12}^2+r_{23}^2+r_{31}^2),
\end{equation}
and
\begin{equation}
r_{ij}=\sqrt{2}R \sin \alpha_k.
\end{equation}
The energy of the three-body system is:
\begin{eqnarray}
E_3&=&\frac{1}{1.5m}\times \nonumber\\
&&[\frac{\hbar^2}{(\sqrt{2}R \sin \alpha_3)^2}+\frac{\hbar^2}{(\sqrt{2}R \sin \alpha_1)^2}+\frac{\hbar^2}{(\sqrt{2}R \sin \alpha_2)^2}]\nonumber\\
&&+V(\sqrt{2}R \sin \alpha_3) +  V(\sqrt{2}R \sin \alpha_1)+ V(\sqrt{2}R \sin \alpha_2)\nonumber\\
&=&\frac{\hbar^2}{ma^2}\times\nonumber\\
&&\Big\{[\frac{1}{(\sqrt{2}\sin \alpha_3)^2}+\frac{1}{(\sqrt{2}\sin \alpha_1)^2}+\frac{1}{(\sqrt{2}\sin \alpha_2)^2}]\frac{1}{1.5x^2}\nonumber\\
&&-ce^{-(x\sqrt{2}\sin \alpha_3-x_s)^2}-ce^{-(x\sqrt{2}\sin \alpha_1-x_s)^2}\nonumber\\
&&-ce^{-(x\sqrt{2}\sin \alpha_2-x_s)^2}\Big\}, \label{e3}
\end{eqnarray}
where $x=R/a$, $\alpha_1$ is in [0, $\pi/2$], $|\frac{1}{3}\pi-\alpha_1|<\alpha_2<\frac{1}{2}\pi-|\frac{1}{6}\pi-\alpha_1|$ and we have the identity
\begin{equation}
\sin\alpha_1^2+\sin\alpha_2^2+\sin\alpha_3^2=\frac{3}{2}.
\end{equation}
Similar to Eq. (\ref{e2}) the systems' dimensions are contained in the term before the curly brackets. Eqs. (\ref{e2}) and (\ref{e3}) can be reduced to a-dimensional forms by defining the universal energies $\varepsilon_2=\frac{E_2}{\hbar^2/ma^2}$ and $\varepsilon_3=\frac{E_3}{\hbar^2/ma^2}$. Some notable cases: three equal hyper-angles (= $\frac{\pi}{4}$) geometrically correspond to an equilater triangle and it is the ground state (TS); one of the hyper-angles equals to $\frac{\pi}{2}$ (the other two hyper-angles can be derived from the constraints above) gives a linear configuration.  Another extreme case is when one of the hyper-angles approaches zero resulting in the monomer+dimer configuration (ES), see fig. 19 in Ref. \cite{Braaten:2006vd}. The last case results in a divergence in Eq. (\ref{e3}) which is the reason for the ES, i.e., $\alpha_i\rightarrow 0$ but $x\sin(\alpha_i)\approx x\alpha_i$ finite, while the other hyper-angles $\alpha_{j, k}\rightarrow \frac{\pi}{3}$.

\begin{figure}
\centering
\begin{tabular}{c}
\includegraphics*[scale=0.4]{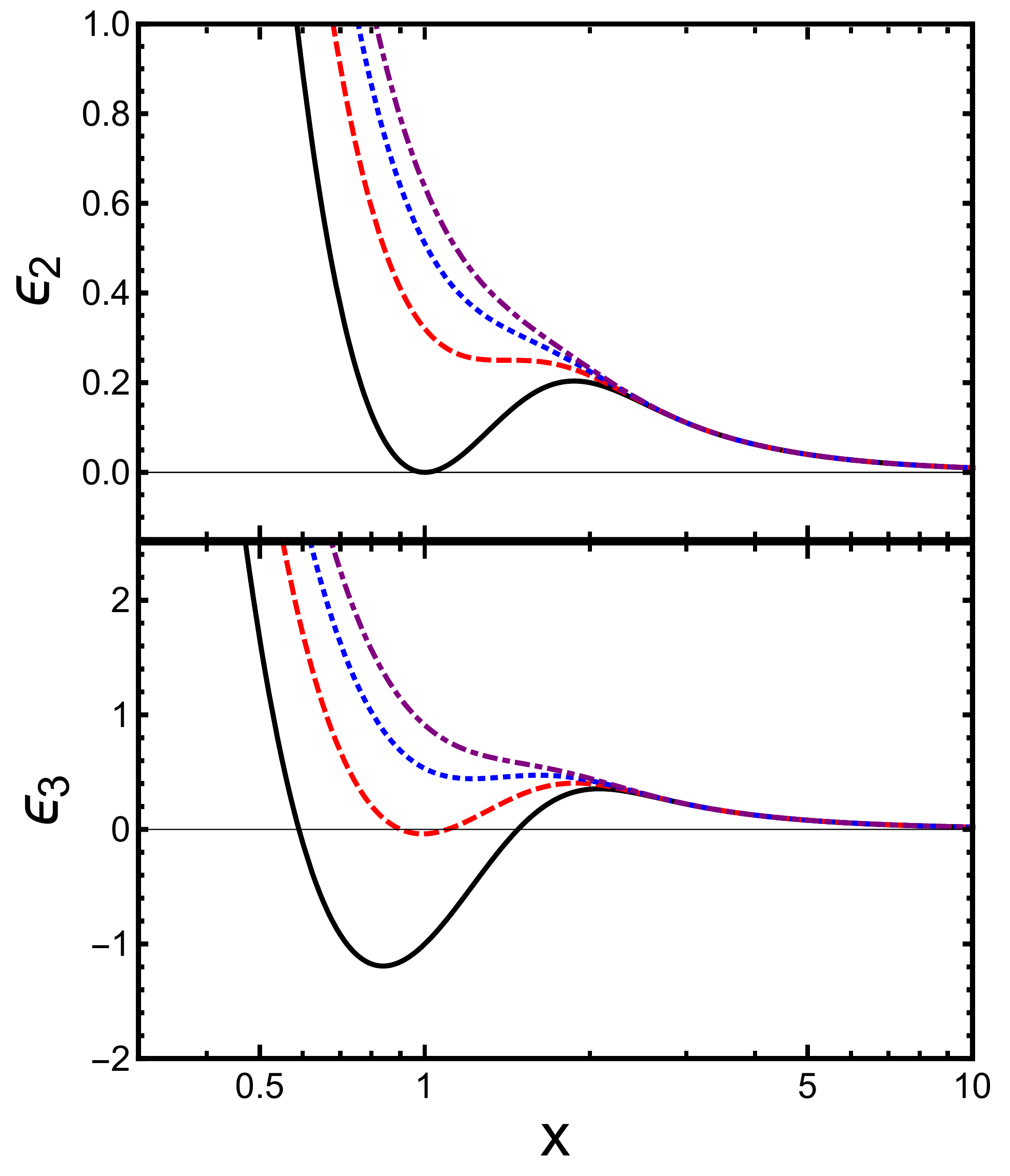}
\end{tabular}
\caption{(Color online) Universal dimer (top) and trimer energy (bottom) vs. scaled hyper-radius x (=$r_{ij}/a$ for two-body system and =R/a for three-body system, see text). The potential strength c is obtained imposing $\varepsilon_2=\varepsilon_2' =0$ (c=2.718, full-black line), $\varepsilon_2'=\varepsilon_2'' =0$, (c=1.847, long-dashed red line), $\varepsilon_2''=\varepsilon_2''' =0$, (c=1.331, short-dashed blue line), $\varepsilon_2'''=\varepsilon_2'''' =0$, (c=0.988, dotted-dashed purple line). Thus bound or resonance states in the trimers disappear when $c<1$.}
\label{figure1}

\end{figure}

In order to derive the Thomas theorem we set $r_0=0$ and interpret $a$ as the potential range \cite{Thomas:1935zz}. In the limit $a\rightarrow 0$ the Gaussian potential in Eq. (\ref{pv}) becomes a $\delta$-function and the term $\frac{\hbar^2}{ma^2}\rightarrow \infty$, thus in order to recover the theorem we need to show that $\varepsilon_3$ has a negative minimum and the minimum of $\varepsilon_2$  equal to zero. There are many other methods to demonstrate the Thomas theorem within our framework and our `short cut' here will be useful for the remaining of the paper. Thus we require that $\varepsilon_2=\varepsilon_2' =0$, i.e., the dimer has a minimum with zero energy. With these conditions we can determine the parameter $c$ in Eq. (\ref{e2}) and derive the trimer energy from Eq. (\ref{e3}). Using symmetry considerations, it is clear that the minimum energy of the three-body system is recovered when the three hyper-angles are equal (TS) and this can be easily demonstrated analytically. In Fig. \ref{figure1} we plot $\varepsilon_2$ (top) and $\varepsilon_3$  (bottom) vs $x$. The dimer is `loosely' bound while the trimer becomes strongly bound and taking $a\rightarrow 0$ results in the diverging $\varepsilon_3$.  We can study under which conditions finite values of the scattering length give still a bound trimer. In Fig. \ref{figure1}  we have plotted the cases where the interaction strength is obtained imposing $\varepsilon_2'=\varepsilon_2'' =0$, (i.e., the energy has a flex)-dashed line, $\varepsilon_2''=\varepsilon_2''' =0$-dotted line and $\varepsilon_2'''=\varepsilon_2'''' =0$-dashed-dotted line. Thus the trimer is loosely bound when the two-body channel displays a flex in energy. A weak two-body attraction still results in a resonance in the trimer.

\begin{figure*}[t]
\centering
\begin{tabular}{c}
\includegraphics*[scale=0.8]{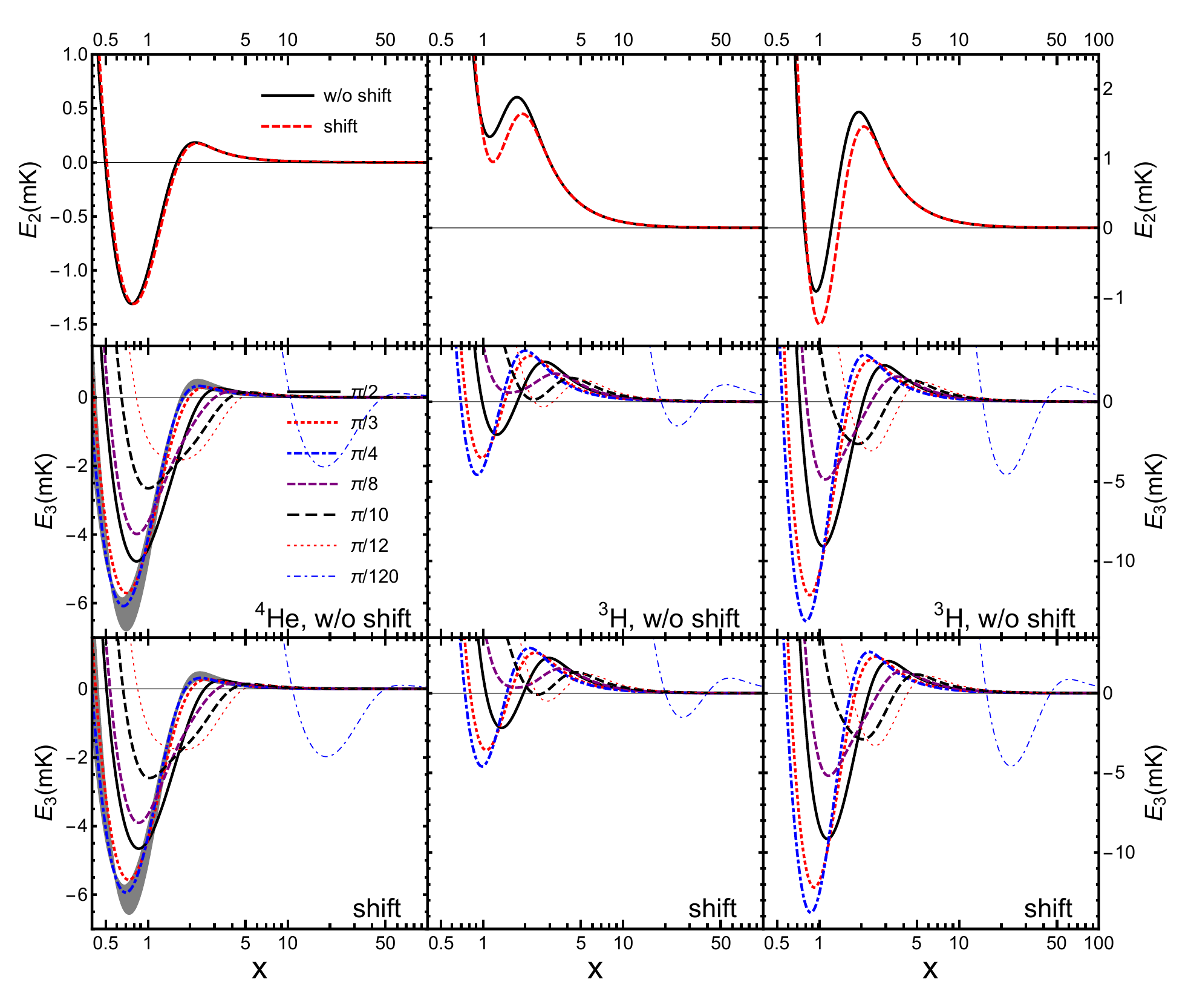}
\end{tabular}
\caption{(Color online) Energy of $^{4}$He (left panels), T (middle panels) atoms fitting the $c$ with TS and T (right panels) atoms fitting the $c$ with ES vs scaled distance $x$. The cases with and without scattering length shifts are clearly marked in the figure.}
\label{figure2}
\end{figure*}

We now discuss some physical systems starting with the $^{4}$He atomic case. This system has been studied at length both experimentally and theoretically \cite{Braaten:2006vd}. The experimental scattering length $a=197_{-34}^{+15}a_0$ \cite{Grisenti:2000zz} where $a_0=5.29177\times10^{-11}$ m is the Bohr radius but the effective range is not known. Theoretical models, which reproduce the large scattering length \cite{Braaten:2006vd, sca1, sca2, sca3, sca4, sca5, sca6, sca7}, give $r_s=13.85a_0$ and such value depends somehow on the chosen interaction. From Eq. (\ref{eq1}) we expect to have at most one ES. The binding energy of the dimer is $E_2=1.1^{+0.3}_{-0.2}$ mK \cite{Braaten:2006vd, Grisenti:2000zz}, which we can use to fit the parameter $c$. Detailed theoretical models give two bound states for the trimer. For instance Motovilov {\it et al.} \cite{sca1} give $E_3^0$=125.8 mK and $E_3^1$=2.28 mK, the highest binding has been confirmed experimentally \cite{be3}. There is a large consensus that $E_3^1$ is an ES  \cite{Braaten:2006vd} and a recent experimental confirmation \cite{Kunitski:2015qth}.  We can build an analogy with the $^{12}$C nucleus, its binding energy is $E_3^0$=(-)92.162 MeV while the Hoyle excited state is at $E_3^1=(7.68-7.65)\times 12$=0.36 MeV. The large difference between the ground state and the excited one is due to the large binding energy of the $\alpha$ particle. In fact, because of the large binding, the first particle that can be emitted (not a $\gamma$) from the excited $^{12}$C is an $\alpha$-particle leaving a $^{8}$Be that subsequently decays into 2$\alpha$ since its g.s. is unbound.  The Hoyle state decays into a monomer+dimer (which later decays again) thus the suggestion that the Hoyle state is an ES \cite{Efimov:1970zz}.  On the other hand, the ground state of $^{12}$C cannot be explained in terms of $\alpha$-clusters but rather of single nucleons degrees of freedom \cite{hoylestateabinitio} and this fact is at the root of its large binding energy thus it is not a TS. Given this preamble, in Fig. \ref{figure2} we plot (left panels) the $E_2$ (top), $E_3$ (middle, $r_s$=0) and $E_3$ (bottom, $r_s=13.85a_0$) vs $x$ for the $^{4}$He atoms. We have fitted the potential parameter to $E_2$ and considered two cases with and without effective range correction. The two limits are very similar since $r_s\ll a$. In the middle panel, we display $E_3$ for $r_s$=0 and we have changed the scattering length within the experimental errors to see changes in the ground energy (shaded region in Fig. \ref{figure2}), corresponding to three equal hyper-angles.  Similar results are obtained for finite effective ranges as expected. As we see from the figure the TS is relatively unbound (6 mK) respect to the theoretical $E_3^0$=125.8 mK, while the ES which can be obtained in the limit $\alpha_i\rightarrow 0$ results in $E_3^1= 1.98_{-0.06}^{+0.2}$ mK, very close to other models \cite{Braaten:2006vd}. Thus, in analogy to the $^{12}$C comparison above, we conclude that the TS does not occur because of the dominance of other degrees of freedom not included in our simple model. Other hyper-angle values, for instance corresponding to the linear configuration, are not the true g.s. (the one with the lowest energy) which is given by the equal sides configuration. If an external field, for instance strong magnets, could influence the interaction and break the symmetry, then one would be able to see other geometrical configurations.

We now turn to the cases where the scattering length is negative. One example is given by the polarized triton (T) atoms whose (triplet) scattering length $a_t=-82.1a_0$, while the effective length is not known and models give $r_t=13.7a_0$ \cite{Braaten:2006vd}. Since the scattering length is negative, there should be no dimer bound state, while the experimental value of the trimer $E_3$=4.59 mK \cite{Braaten:2006vd, he3atom}. Since we cannot fit the $c$-value to $E_2$, we decided to fit $E_3$ instead assuming that its experimental value corresponds to the TS. In the middle panels of figure \ref{figure2} we report the results obtained under this (strong) assumption. The cases with zero (middle) and finite (bottom) effective range are given as well. Since we fix the parameter to $E_3$ the two results are not much different while some difference can be noticed in $E_2$, which now becomes a model prediction for the dimer suggesting a resonant state.  If our assumption that the TS is the experimental $E_3$ then we expect an ES at $E_3^1$=1.53 mK. On the other hand, if the TS is hindered by other degrees of freedom, in analogy with the $^{4}$He and $^{12}$C above, then we should fit the parameter $c$ to the ES. In the right panels of Fig. \ref{figure2}, we plot the results where we fit the parameter $c$ assuming that the ES  energy is equal to  the experimental $E_3$. As we see from the figure, the TS becomes largely bound as well as the dimer state in contrast with the negative value of the scattering length. Thus our approach suggests that there might be an ES around 1.53 mK.  
Our ansatz can be tested experimentally, but what we want to point out as a consequence of negative scattering lengths is the appearance of unbound levels when $\alpha_i\approx \frac{\pi}{8}-\frac{\pi}{10}$, figure \ref{figure2}. 

\begin{figure}
\centering
\begin{tabular}{c}
\includegraphics*[scale=0.4]{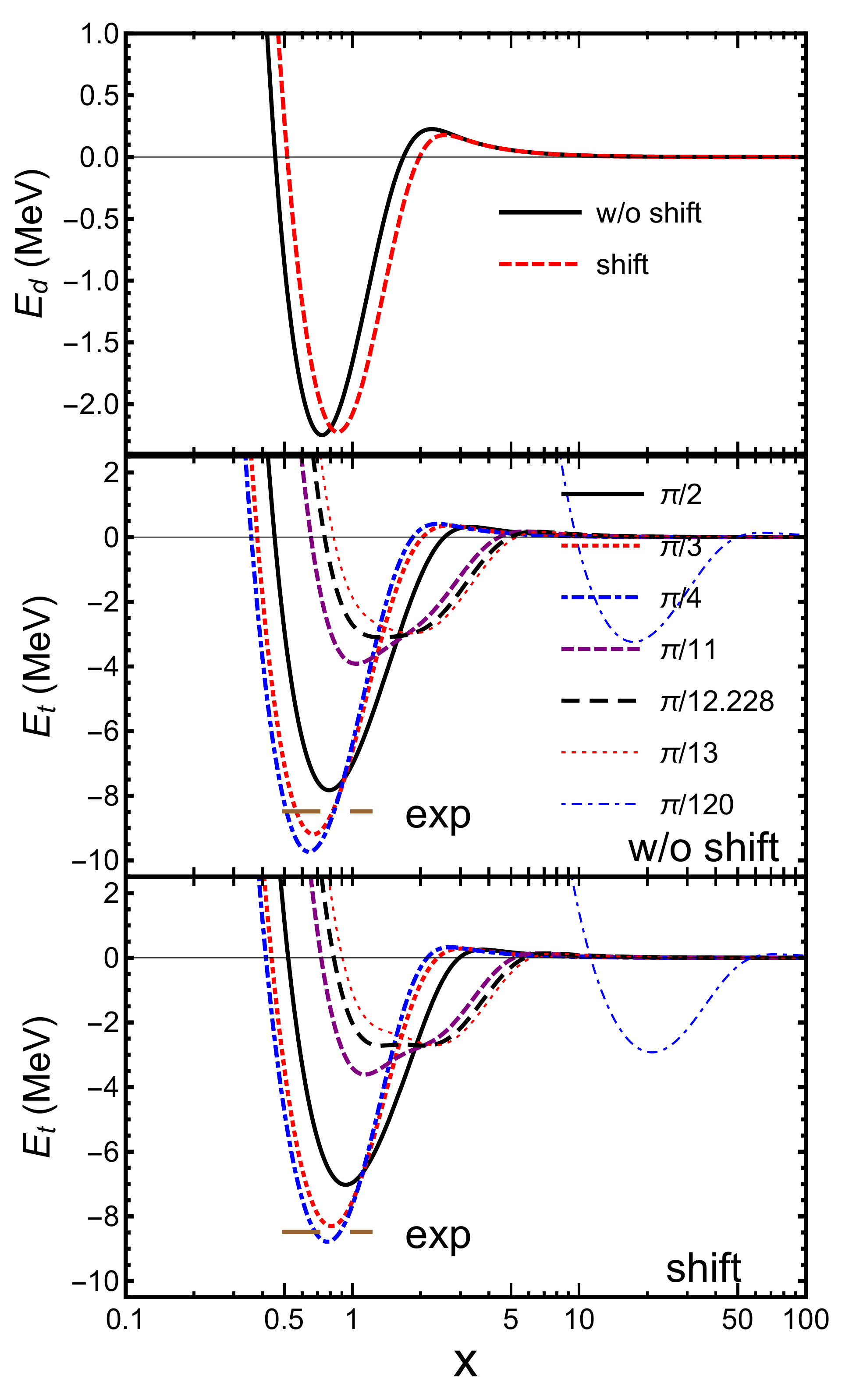}
\end{tabular}
\caption{(Color online) Deuterium (top panel) and triton binding energies without (middle panel) and with shift (bottom panel)  in the two-body potential. Different hyper-angle values are indicated. Two equal minima can be seen in the bottom panel for $\alpha_1=\pi/12.228$, indicating the transition from TS to ES.}
\label{figure3}
\end{figure}

For t nuclei all the physical quantities needed to test our model are well known and reported in the introduction.  A complication might be due to the fact that the interaction between protons and neutrons and neutrons-neutrons with different spins might not be the same. We will assume the same potential for all non-identical Fermions (the two neutrons must have different spins because of the Pauli blocking). The small difference between our results and experiments will be discussed in more detail in a following paper. In figure \ref{figure3}, we plot the d and t energies vs $x$ with and without effective range correction. The parameter $c$ was fitted to the binding energy of the deuterium. We see a very good agreement with $r_s$ given by the experimental value. Since we do not expect the quarks degrees of freedom to play an important role, we conclude that the trimer ground state is indeed a TS.  Small hyper-angles give an ES at about 7 MeV excitation energy in agreement with the experimental results \cite{Rogachev:2003mv, russ}, which still await further experimental confirmation. Another interesting feature arises and can be clearly seen in the figure \ref{figure3}. When the hyper-angle approaches a critical value $\alpha_{ic}=\frac{\pi}{12.228}$ two equal minima appear. We interpret such critical value as the border between TS, i.e., larger hyper-angles result in one absolute minimum, and the ES, i.e., smaller $\alpha_i$ result in an excited level which scales as the product $x\alpha_i$.  The critical $\alpha_i$ appears when the $r_s$ is close to $a$, compare figures \ref{figure2} and \ref{figure3}, and its shape resembles an isotropic-nematic first order phase transition. The question arises: when the $r_s$ is negligible does the phase transition become second order?

\begin{figure}
\centering
\begin{tabular}{c}
\includegraphics*[scale=0.4]{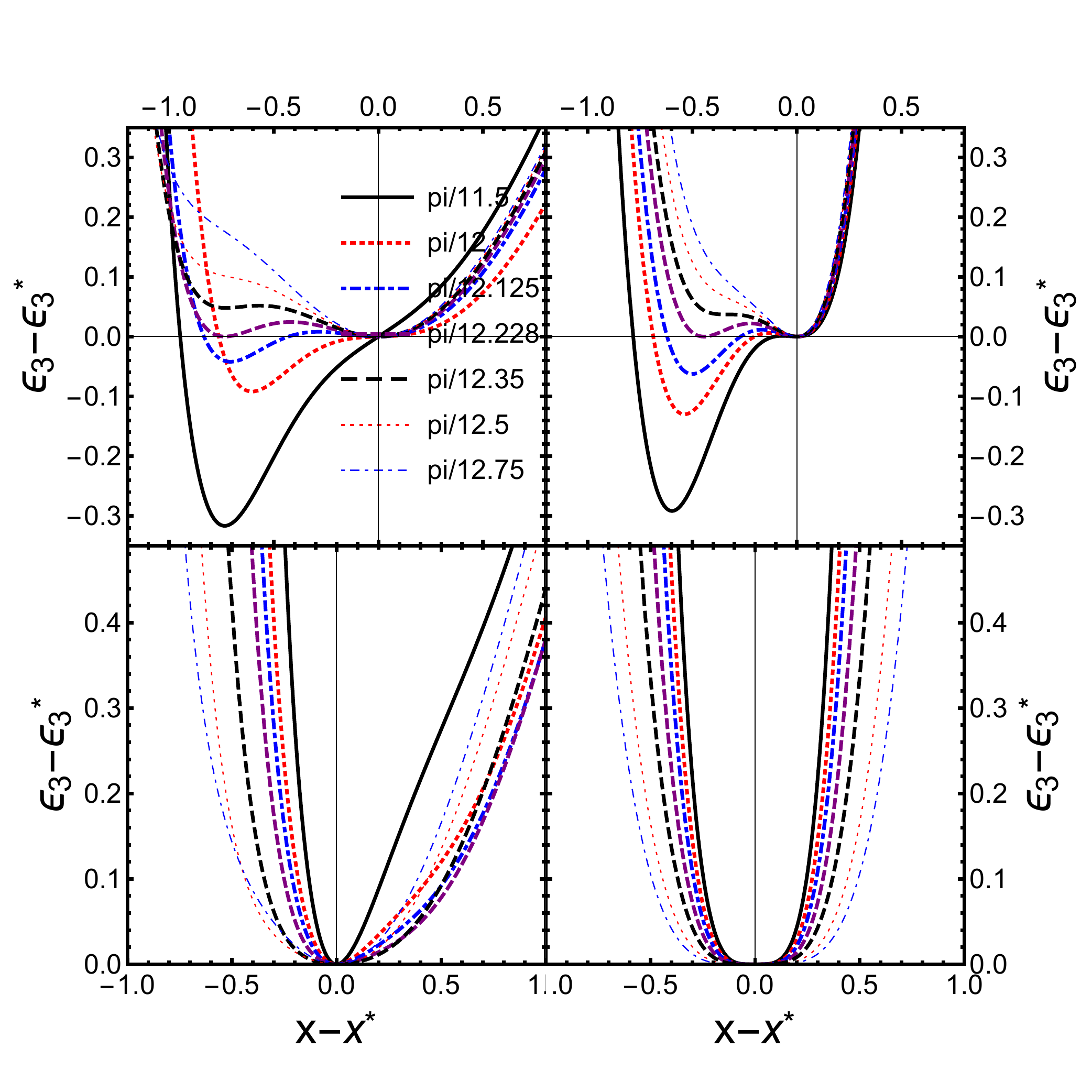}
\end{tabular}
\caption{(Color online) (Top panels) A-dimensional energy opportunely shifted vs $x-x^*$ for cases where the ratio $\frac{r_s}{a}$ is not small (left) and the results using the Landau's (free) energy, Eq. (\ref{lfe}) for an opportune choice of the fitting parameters (right); (Bottom panels) Same as the top panels, for case where the ratio $\frac{r_s}{a}$=0 (left) and $e=0$ and $\alpha_{i^*}=\alpha_{ic}$ in Landau's (free) energy, Eq. (\ref{lfe}).}
\label{figure4}
\end{figure}

Following Landau's theory of phase transitions we assume that the (free) energy has a singular part, which can be written as:
\begin{equation}
\varepsilon_3=\varepsilon_3^*+d(\alpha_i-\alpha_{i^*})\frac{(x-x^*)^2}{2}+e\frac{(x-x^*)^3}{3}+f\frac{(x-x^*)^4}{4}+\cdots \label{lfe}
\end{equation}
where we have neglected any external field \cite{huang}. $d$, $e$ and $f$ are fitting parameters. We have assumed that $\alpha_i$ and $x$ play the roles of the control and order parameters respectively.  In figure \ref{figure4} (top panels) we plot the a-dimensional energy opportunely shifted vs $x-x^*$ for cases where the ratio $\frac{r_s}{a}$ is not small, see Eq. (\ref{lfe}). In the right panel we have plotted the results using the Landau's (free) energy, Eq. (\ref{lfe}), for an opportune choice of the fitting parameters. 

The odd order term in Eq. (\ref{lfe}) is crucial and breaks the symmetry $(x-x^*)\rightarrow-(x-x^*)$ similar to a nematic first order phase transition \cite{revlandau}. The phase transition is due to a change of shape, in fact for $\alpha_i=\alpha_{ic}$ the two minima corresponding to two different distances between the monomer and dimer as given by $x_{jk}=\sqrt{2}x \sin(\alpha_i)$, it also signals the point where we have a transition from the TS (dominated by the absolute minimum for equal energies) and the ES where the energy scales with $x\sin(\alpha_i)$. Taking the ratio $\frac{r_s}{a}\rightarrow 0$, the two minima disappear (for $\frac{r_s}{a}$=0.12) and now we can reproduce the reduced energy vs x with the Landau's form with e=0 and $\alpha_i^*=\alpha_{ic}$, Eq. (\ref{lfe}). 

In conclusion, in this work we have discussed a simple model based on the Bohr atom and hyper-coordinates to study the transition from Thomas states to Efimov states. We first recovered the Thomas theorem. We have shown the the t nuclei have both the TS and ES if existing experiments are confirmed, while in $^{4}$He and $^{3}$He atoms (as well as $^{12}$C nucleus) the TS might disappear because other degrees of freedom (for instance nucleons vs $\alpha$-clusters in $^{12}$C) become dominant giving rise to a deeper minimum. We feel confident that the transparent features discussed here will be of great help in guiding future experimental and theoretical investigations.

\section*{Acknowledgments}
This work was supported by the US Department of Energy under Grant No. DE-FG02-93ER40773, NNSA DE-NA0003841 (CENTAUR).

\end{document}